\documentclass[a4paper,12pt]{llncs}

\usepackage{latexsym}
\usepackage{graphicx}
\usepackage{amsmath, amssymb}
\usepackage{a4}
\usepackage[ruled,vlined]{algorithm2e}
\usepackage{authblk}
\usepackage{xcolor}
\usepackage{tikz}

\usepackage{hyperref}
\usepackage{verbatim}
\renewcommand{\thefigure}{\arabic{figure}}
\hypersetup{
  colorlinks=true,
  linkcolor=blue,
  citecolor=black,
  urlcolor=blue
}

\makeatletter
\renewcommand{\fnum@figure}{Fig. \thefigure}
\makeatother

\newcommand{\dusm}{\ensuremath{\delta_{\textit{USM}}}}
\newcommand{\ddiscretePO}{\ensuremath{\delta_{\textit{d}}}}

\newcommand{\continuousPO}{{\delta_c}}
\newcommand{\leftI}{I^-}
\newcommand{\rightI}{I^+}

\newcommand{\wolf}[1]{\textcolor{purple}{Wolf: \textbf{#1}}}
\newcommand{\maike}[1]{\textcolor{cyan}{Maike: \textbf{#1}}}
\newcommand{\jona}[1]{\textcolor{blue}{Jona: \textbf{#1}}}

\newcommand{\takeout}[1]{}
\renewcommand{\wolf}[1]{} \renewcommand{\maike}[1]{} \renewcommand{\jona}[1]{} 

\title{Fr\'echet Distance of Partially Ordered Point Sets}

\author{Maike Buchin \inst{1}\orcidID{0000-0002-3446-4343} \and Jona Handzsuj \inst{2}\orcidID{0009-0009-5566-4887} \and Wolf Ki\ss ler\inst{1}\orcidID{0009-0004-5609-5739}}

\institute{Faculty of Computer Science, Ruhr University Bochum \\\texttt{\{maike.buchin, wolf.kissler\}@ruhr-uni-bochum.de}
\and FernUniversität Hagen, Germany \\\texttt{jona.handzsuj@fernuni-hagen.de}}

\begin{document}

\maketitle

\begin{abstract}
We introduce discrete and continuous Fréchet distances for partially ordered point sets.
Each linear extension of a partial order defines one possible polygonal curve, and the distance between two input posets is the minimum Fréchet distance over all such curves. 
This includes the classical Fréchet distance for ordered curves and the unordered point-set variant as special cases. These problems are NP-hard and we give exact exponential dynamic programming algorithms for computing both distances.
\end{abstract}


\section{Introduction}

The Fréchet distance is a 
similarity measure for 
curves.
Intuitively, it measures how closely two curves can be traversed at the same time, under the constraint that both traversals move continuously and monotonically from start to end.
In contrast to purely set-based distances, such as the Hausdorff distance, the Fréchet distance respects the sequential nature of the input.
This makes it particularly suitable for comparing trajectories, time series, and other ordered geometric data.

In many applications, however, the input is not naturally given as a pair of fully ordered point sequences.
Instead, the available information may specify only partial precedence constraints between points.
Such partial orders capture settings in which some ordering information is known, while other relative positions remain ambiguous. 
For example, trajectory points may have uncertain time stamps: if the possible
time interval of a point $x$ lies completely before that of a point $y$, we
impose $x \prec y$, while overlapping time intervals leave the two points incomparable. 
Trajectory points may also be missing a time stamp due to incomplete data. 
Partial orders also arise from unsynchronized sensors, where the
order within each sensor stream is known but measurements from different
streams cannot always be globally ordered.
This motivates the study of curve similarity measures that can account for incomplete or uncertain ordering information.

We study a natural generalization of the Fréchet distance to partially ordered point sets.
The basic idea is to regard every linear extension of a partial order as one possible curve consistent with the available information.
The distance between two partially ordered point sets is then the minimum Fréchet distance over all pairs of curves induced by linear extensions of their respective partial orders.
Thus, the model interpolates between the classical Fréchet distance of polygonal curves, where the order is completely fixed, and the unordered point-set variant, where all orders are allowed.

\subsection{Preliminaries and problem definition}
\label{sec:prelim}

For ease of presentation, we state all definitions in $\mathbb{R}^2$. They extend directly to $\mathbb{R}^d$ for every $d\in \mathbb N$.
A \emph{polygonal curve} with \emph{vertices} $p_1,\dots,p_n$ is the map $P\colon[1,n]\to\mathbb{R}^2$ with $P(i)=p_i$ that is affine on each interval $[i,i+1]$, for $i=1,\dots,n-1$.
For polygonal curves $P\colon[1,n]\to\mathbb{R}^2$ and $Q\colon[1,m]\to\mathbb{R}^2$, the \emph{Fréchet distance} is
$$
\delta_{F}(P,Q)=\inf_{\alpha,\beta}\ \max_{t\in[0,1]}
\bigl\lVert P(\alpha(t))-Q(\beta(t))\bigr\rVert ,
$$
where $\alpha\colon[0,1]\to[1,n]$ and $\beta\colon[0,1]\to[1,m]$ range over all continuous non-decreasing surjections, called \emph{reparametrizations} of $P$ and $Q$.

The \emph{discrete Fréchet distance} restricts distance comparisons to the vertices.
A \emph{coupling} between $P$ and $Q$ is a sequence $(a_1,b_1),\dots,(a_K,b_K)$ of index pairs with $(a_1,b_1)=(1,1)$, $(a_K,b_K)=(n,m)$ and $(a_{i+1},b_{i+1})-(a_i,b_i)\in\{(0,1),(1,0),(1,1)\}$ for all $i<K$.
Its \emph{cost} is $\max_i\lVert p_{a_i}-q_{b_i}\rVert$, and the discrete Fréchet distance $\delta_{dF}(P,Q)$ is the minimum cost of a coupling between $P$ and $Q$.

A \emph{partially ordered point set} is a finite point set $S\subset\mathbb{R}^2$ together with a strict partial order $\prec_S$ on $S$, where $x\prec_S y$ indicates that $x$ must be visited before $y$. 
We usually write $S$ for $(S,\prec_S)$.
A subset $X'\subseteq X$ of a partially ordered point set $X$ is a \emph{lower set} if $x\in X'$ and $y\prec_X x$ imply $y\in X'$.
We write $\mathcal J(X)$ for the family of all lower sets of $X$, and $\operatorname{Max}(X')=\{x\in X'\mid\text{there is no }y\in X'\text{ with }x\prec_X y\}$ for the set of maximal elements of $X'$.
The \emph{width} $w_X$ of $X$ is the maximum size of an antichain of $\prec_X$; in particular $|\operatorname{Max}(X')|\le w_X$ for every $X'\subseteq X$.
A total order $\pi$ of $S$ is a \emph{linear extension} of $\prec_S$ if $x\prec_S y$ implies that $x$ precedes $y$ in $\pi$. 
It induces the polygonal curve $c_\pi(S)$ that visits every point of $S$ exactly once in the order given by $\pi$.
The curves of this form are exactly the curves \emph{induced} by $S$.
We extend both Fréchet distances to partially ordered point sets $S$ and $R$ by minimizing over all induced curves,
$$
\delta_{dF}(S,R):=\min_{\pi,\rho}\delta_{dF}\bigl(c_\pi(S),c_\rho(R)\bigr),
\qquad
\delta_{F}(S,R):=\min_{\pi,\rho}\delta_{F}\bigl(c_\pi(S),c_\rho(R)\bigr),
$$
where $\pi$ and $\rho$ range over the linear extensions of $\prec_S$ and $\prec_R$.
This reduces to the classical (discrete) Fréchet distance when both partial orders are total, and to the point set variant of Buchin and Kilgus~\cite{BUCHIN2022} when both are empty.

\subsection{Related work}
 
Alt and Godau~\cite{AltGodau1995} gave the classical free-space diagram
algorithm, which for two polygonal curves of complexities $n$ and $m$ decides in
$O(nm)$ time whether their continuous Fréchet distance is at most a given
threshold $\varepsilon$ and computes the distance in $O(nm\log(nm))$ time.
The discrete Fréchet distance was introduced by Eiter and
Mannila~\cite{EiterMannila1994}, who gave an $O(nm)$-time dynamic programming algorithm.
Both algorithms crucially use that the vertex orders are fixed.
 
In curve-to-point-set matching, one input is a polygonal curve $P$ and the other
a point set $S$, and the goal is to find a polygonal curve on points of $S$ close
to $P$. 
The variants require either all points of $S$ or only a subset, and
either allow or forbid repeated visits.
For the discrete Fréchet distance, Wylie and Zhu~\cite{WylieZhu2013} showed that
the non-unique variants are solvable in polynomial time while the unique variants
are NP-complete, and that this also implies hardness for the continuous Fréchet
distance. They did not give algorithms for the unique variants.
 
The case where both input posets are unordered was studied by Buchin and
Kilgus~\cite{BUCHIN2022}.
They showed that the discrete Fréchet distance of two point sets equals their
Hausdorff distance, so the difficulty lies not in computing the value but in
constructing representative curves that realize it. 
Shortest realizing curves
need not be simple, and they studied the construction of short and simple
representatives.
For the continuous variant they showed NP-hardness, even for constant-factor
approximation, and gave an exponential-time algorithm whose running time depends
on geometric properties of the instance.

A related but different setting is when the points have spatial but not temporal uncertainty. The problem of computing the Fréchet distance of uncertain curves has been studied for both the discrete~\cite{AhnKSSV2012} and continuous variant~\cite{BuchinFLPRR2023}. 

\subsection{Contribution}

We present dynamic-programming algorithms for computing Fréchet distances between partially ordered point sets.
The corresponding decision problems are NP-hard, since they contain the point-set-to-curve variants studied by Wylie and Zhu~\cite{WylieZhu2013} as special cases, so exact polynomial-time algorithms are unlikely.
Our algorithms follow the Bellman--Held--Karp approach for the traveling salesperson problem~\cite{Bellman1962,HeldKarp1962} and replace the naive enumeration of all linear extensions by subset-based dynamic programming algorithms.

As an introduction to the core ideas, we compute the discrete Fréchet distance between an unordered point set $S$ with $|S|=n$ and a totally ordered point set $P$ with $m$ vertices in $\mathcal O(nm2^n)$ time.

Extending the approach to two arbitrary partially ordered point sets $S$ and $R$ of sizes $n$ and $m$ gives a dynamic programming algorithm that is indexed by pairs of lower sets and computes $\delta_{dF}(S,R)$ in $\mathcal O(nm2^{n+m})$ time.

For the continuous Fréchet distance we combine this approach with the free-space reachability framework of Alt and Godau~\cite{AltGodau1995}.
Here the two traversals may meet in the interior of an edge, so a state additionally records how far along its current edge each traversal already is, and stores the earliest position on that edge that can be reached with leash length~$\varepsilon$.
For a fixed $\varepsilon$, this decides whether $\delta_F(S,R)\le\varepsilon$ in $\mathcal O(n^2m^22^{n+m})$ time, and binary search over the standard set of critical values yields the exact value in $\mathcal O(n^2m^22^{n+m}\log(nm))$ time.

All three running times are single exponential, and the exponential factor counts pairs of lower sets.
A state of the algorithms for two partially ordered point sets is a pair of lower sets together with an endpoint on each side, and these endpoints range over antichains, so there are at most $w_Sw_R\,|\mathcal J(S)|\,|\mathcal J(R)|$ states in the discrete and $\mathcal O\bigl(w_S^2w_R^2\,|\mathcal J(S)|\,|\mathcal J(R)|\bigr)$ state evaluations in the continuous case.
The bounds above are the worst case, attained for empty orders, whereas for total orders $w_S=w_R=1$, $|\mathcal J(S)|=n+1$ and $|\mathcal J(R)|=m+1$, leaving $\mathcal O(nm)$ states, on which our dynamic programming algorithms specialize to the classical ones of Eiter and Mannila~\cite{EiterMannila1994} and of Alt and Godau~\cite{AltGodau1995}.
More ordering information in the input thus translates directly into a faster computation.
The single-exponential dependence cannot be improved much: 3-SAT with $c$ clauses reduces to our problem with $n+m=\mathcal O(c)$. 
As refutations of the Exponential Time Hypothesis (ETH) have been announced~\cite{Ope26,Ant26}, we use the weaker Satisfiability Hypothesis (SH)~\cite{StearnsHunt1990}, under which the exponent $n+m$ cannot be improved to $(n+m)^{1-\gamma}$ for any constant $\gamma>0$.
\section{Algorithms}

Section~\ref{sec:Alg1} treats an unordered point set and a curve,
Sections~\ref{sec:Alg2} and~\ref{sec:Alg3} two posets under
the discrete and the continuous Fréchet distance. We write ${d(x,y)=\|x-y\|}$ for the Euclidean distance and assume the real RAM model.

\subsection{Discrete Fréchet distance of a point set and a curve}
\label{sec:Alg1}

\begin{figure}[t]
    \centering
    \includegraphics[width=\textwidth]{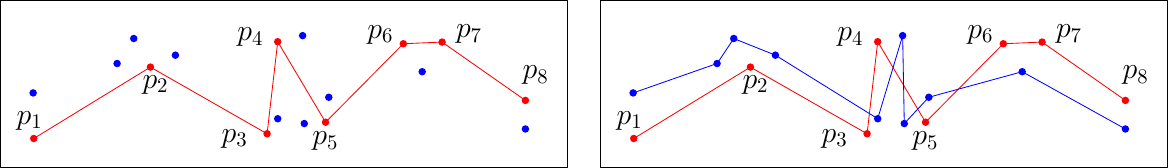}
    \caption{An instance of the point-set-to-curve problem (left) and an optimal curve for it (right). The curve $P$ is shown in red, and an induced curve on the unordered point set $S$ is shown in blue.}
    \label{fig:example}
\end{figure}

Let $S=\{s_1,\ldots,s_n\}$ be an unordered point set, that is, $\prec_S$ is empty, and let $P=(p_1,\ldots,p_m)$ be a polygonal curve, viewed as a totally ordered point set.
We write $P_k=(p_1,\ldots,p_k)$ for its prefix of length $k$.
The task is to compute $\delta_{dF}(S,P)=\min_\pi\delta_{dF}(c_\pi(S),P)$, where $\pi$ ranges over all linear extensions of $\prec_S$, hence over all orderings of $S$; for an example, see Figure~\ref{fig:example}.
This is the all-points variant of the unique set-chain matching problem of Wylie and Zhu~\cite{WylieZhu2013}, whose decision version is NP-hard, and the naive algorithm, which runs the standard $\mathcal O(nm)$ dynamic programming algorithm for each of the $n!$ orderings of $S$, takes $\mathcal O(n!\,nm)$ time.

We use table entries $\dusm(s_i,S',k)$ with $S'\subseteq S$, $s_i\in S'$ and $k\in\{1,\ldots,m\}$, where $\dusm(s_i,S',k)$ is intended to store the minimum discrete Fréchet distance between $P_k$ and a curve induced by $S'$ whose last vertex is $s_i$.
We further maintain the auxiliary table $M(S',k):=\min_{s_i\in S'}\dusm(s_i,S',k)$, which stores the best value of the same subproblem when the endpoint is not fixed and thereby avoids recomputing the minimum over all possible predecessor endpoints.
For $|S'|>1$ and $k>1$, the recurrence is
$$
\dusm(s_i, S', k)
    =
    \max\left\{
        d(s_i,p_k),
        \min\left\{
        \begin{aligned}
            &M(S' \setminus \{s_i\}, k-1),\\
            &M(S' \setminus \{s_i\}, k),\\
            &\dusm(s_i, S', k-1)
        \end{aligned}
        \right\}
    \right\}.
$$
Its three inner terms correspond to the three possible predecessors $(s_j,p_{k-1})$, $(s_j,p_k)$ with $s_j\in S'\setminus\{s_i\}$, and $(s_i,p_{k-1})$ of the pair $(s_i,p_k)$ in a discrete Fréchet coupling.
In the first two cases the auxiliary table suffices, since after appending $(s_i,p_k)$ the later subproblem only depends on the fact that the curve ends at $s_i$, not on which point $s_j$ came before.
The boundary cases $k=1$ and $S'=\{s_i\}$ are
$$
\dusm(s_i,S',1)=\max_{s_j\in S'}d(s_j,p_1)
\qquad\text{and}\qquad
\dusm(s_i,\{s_i\},k)=\max_{1\le j\le k}d(s_i,p_j),
$$
for all non-empty $S'\subseteq S$, $s_i\in S'$ and $k\in\{1,\ldots,m\}$.
The dynamic programming algorithm first computes these values together with the corresponding values of $M$, and then the remaining entries in increasing order of $k$ and, for equal $k$, of $|S'|$, setting $M(S',k)=\min_{s_i\in S'}\dusm(s_i,S',k)$ as soon as all entries for a pair $(S',k)$ are known.
It finally returns $M(S,m)$.
In this order, the entries for a fixed $k$ depend only on entries with the same $k$ and a strictly smaller set, or on entries for $k-1$, so for computing $M(S,m)$ it suffices to keep two consecutive values of $k$ in memory.
Storing the predecessor choices of all entries instead allows the reconstruction of a linear extension $\pi$ and an induced curve $c_\pi(S)$ that minimize $\delta_{dF}(c_\pi(S), P)$.

\begin{theorem} \label{thm:Algo1}
    Given an unordered point set $S$ of size $|S|=n$ and a polygonal curve $P$ with $m$ vertices, the value $M(S,m)$ equals $\delta_{dF}(S,P)$ and can be computed in $\mathcal{O}(nm2^n)$ time and $\mathcal{O}(n2^n)$ space.
    A linear extension and thus an induced curve realizing this value can be computed in $\mathcal{O}(nm2^n)$ time and space.
\end{theorem}

The proof is by induction on $|S'|+k$; we defer it to Appendix~\ref{app:proofAlgo1}, since most of its argument reappears in the proof of Theorem~\ref{thm:Algo2}.
\subsection{Discrete Fréchet distance of partially ordered point sets}
\label{sec:Alg2}

We now extend the dynamic programming algorithm from the previous subsection to the case where
both input posets are partially ordered point sets.
Let $S=\{s_1,\ldots,s_n\}$ and $R=\{r_1,\ldots,r_m\}$ have partial orders
$\prec_S$ and $\prec_R$, respectively.
There are two differences from the previous algorithm.
Both induced curves are now chosen by the algorithm, and both partial orders are
now arbitrary rather than empty or total.
A state therefore has to record a prefix of an induced curve on each of the two
sides, together with the last point of each of these prefixes, and the algorithm
has to determine which sets of points form such a prefix and which points such a
prefix can end in.

Two facts from Section~\ref{sec:prelim} determine which states are needed.
First, the points visited by a prefix of an induced curve on $S$ contain all
predecessors of each of their points, so they form a lower set, and conversely
every lower set of $S$ is visited by the prefix of some induced curve.
Second, the point at which such a prefix ends is the last element of a linear
extension of that lower set, hence maximal in it, and every maximal element
occurs as such an endpoint~\cite{Stanley2011}.
The same holds for $R$.
We therefore index states by a pair of non-empty lower sets $S'\in\mathcal J(S)$
and $R'\in\mathcal J(R)$ together with endpoints $s_i\in\operatorname{Max}(S')$
and $r_\ell\in\operatorname{Max}(R')$.
All remaining states are infeasible and are never stored.

For such a state we use the table entry ${\ddiscretePO}(s_i,S',r_\ell,R')$.
The entry corresponds to the subproblem in which the induced curve on $S'$ ends
at $s_i$, and the induced curve on $R'$ ends at $r_\ell$.
More precisely, ${\ddiscretePO}(s_i,S',r_\ell,R')$ is intended to store the
minimum discrete Fréchet distance between a curve induced by $S'$ whose last
vertex is $s_i$ and a curve induced by $R'$ whose last vertex is $r_\ell$.

As in the previous subsection, we store auxiliary minima over endpoint choices.
For all non-empty lower sets $S'\in\mathcal J(S)$ and $R'\in\mathcal J(R)$, and
for $s_i\in\operatorname{Max}(S')$ and $r_\ell\in\operatorname{Max}(R')$, define
$$
\begin{aligned}
M_{SR}(S',R')
    &:=
    \min_{\substack{s_j\in \operatorname{Max}(S')\\ r_h\in \operatorname{Max}(R')}}
    {\ddiscretePO}(s_j,S',r_h,R'),
    \\
M_S(S',r_\ell,R')
    &:=
    \min_{s_j\in \operatorname{Max}(S')}
    {\ddiscretePO}(s_j,S',r_\ell,R'),
    \\
M_R(s_i,S',R')
    &:=
    \min_{r_h\in \operatorname{Max}(R')}
    {\ddiscretePO}(s_i,S',r_h,R').
\end{aligned}
$$

The three tables store the best value of the same subproblem when one or both
endpoints are not fixed.
As before, this avoids recomputing the minimum over all possible predecessor
endpoints.

For $|S'|>1$ and $|R'|>1$, the recurrence is
$$
{\ddiscretePO}(s_i,S',r_\ell,R')
=
\max\left\{
    d(s_i,r_\ell),
    \min\left\{
    \begin{aligned}
        &M_{SR}(S'\setminus\{s_i\},R'\setminus\{r_\ell\}),\\
        &M_S(S'\setminus\{s_i\},r_\ell,R'),\\
        &M_R(s_i,S',R'\setminus\{r_\ell\})
    \end{aligned}
    \right\}
\right\}.
$$

The three terms inside the inner minimum correspond to the three possible types
of predecessor in a discrete Fréchet coupling.
The pair preceding $(s_i, r_\ell)$ is of one of the following forms: $(s_j,r_h)$ for some
$s_j\in S'\setminus\{s_i\}$ and $r_h\in R'\setminus\{r_\ell\}$, $(s_j,r_\ell)$
for some $s_j\in S'\setminus\{s_i\}$, or $(s_i,r_h)$ for some
$r_h\in R'\setminus\{r_\ell\}$.
The auxiliary tables may be used in all three cases, since after appending the pair $(s_i,r_\ell)$ the earlier subproblem only depends on the two point
sets and on the endpoints that are still fixed.
Note that the arguments of the three terms are again states of the form
described above.
Removing a maximal element from a lower set leaves a lower set, and the
predecessor endpoints range exactly over the maximal elements of the resulting
sets.

The boundary cases, with $|S'|=1$ or $|R'|=1$, are defined by
$$
\begin{aligned}
{\ddiscretePO}(s_i,\{s_i\},r_\ell,R')
    &=
    \max_{r_h\in R'} d(s_i,r_h),
    &&
    \forall\text{ non-empty } R'\in\mathcal J(R)
    \text{ and } r_\ell\in \operatorname{Max}(R'),
    \\
{\ddiscretePO}(s_i,S',r_\ell,\{r_\ell\})
    &=
    \max_{s_j\in S'} d(s_j,r_\ell),
    &&
    \forall\text{ non-empty } S'\in\mathcal J(S)
    \text{ and } s_i\in \operatorname{Max}(S').
\end{aligned}
$$

A dynamic programming algorithm incrementally fills the table entries.
It first computes the boundary cases and the corresponding values of the
auxiliary tables.
It then computes the remaining values in increasing order of $|S'|+|R'|$.
Whenever all entries ${\ddiscretePO}(s_i,S',r_\ell,R')$ for a fixed pair
$(S',R')$ have been computed, the algorithm computes the corresponding values of
$M_{SR}$, $M_S$, and $M_R$.
After the table is filled, the algorithm returns $M_{SR}(S,R)$.
Next we prove correctness and running time of this algorithm.

\begin{theorem}
\label{thm:Algo2}
    Given partially ordered point sets $S$ and $R$ of sizes $n$ and $m$, the
    value $M_{SR}(S,R)$ equals $\delta_{dF}(S,R)$ and can be computed in
    $\mathcal{O}(nm2^{n+m})$ time and $\mathcal{O}(nm2^{n+m})$ space.
\end{theorem}

\begin{proof}[sketch]
For a state as above, let $\Delta(s_i,S',r_\ell,R')$ be the minimum discrete
Fréchet distance between a curve induced by $S'$ ending in $s_i$ and a curve
induced by $R'$ ending in $r_\ell$.
Since $S'$ and $R'$ are lower sets with maximal endpoints, such curves exist and
$\Delta$ is finite.
We show ${\ddiscretePO}=\Delta$ by induction on $|S'|+|R'|$; then $M_{SR}$, $M_S$
and $M_R$ are correct as minima of correct values.
In the base cases the order on one of the two sets is irrelevant: if $|S'|=1$,
every point of $R'$ is matched to $s_i$, and symmetrically for $|R'|=1$.
Otherwise, the last pair of an optimal coupling is $(s_i,r_\ell)$, and deleting it
leaves a coupling whose last pair is $(s_j,r_h)$, $(s_j,r_\ell)$ or $(s_i,r_h)$,
depending on which indices decrease.
Where they occur, $s_j$ and $r_h$ precede $s_i$ and $r_\ell$ in the linear
extensions, so they are maximal in the lower sets $S'\setminus\{s_i\}$ and
$R'\setminus\{r_\ell\}$. The corresponding entries are therefore states, and by the
induction hypothesis the best predecessor of each type has the corresponding
auxiliary value.
Conversely, each such predecessor coupling extends by $(s_i,r_\ell)$, since
appending a maximal element to a linear extension of the remaining points again
yields a linear extension.
As this turns a value $L$ into $\max\{L,d(s_i,r_\ell)\}$, independently of the
predecessor, the recurrence is exactly the optimum, and minimizing over the last
vertices gives $\delta_{dF}(S,R)=M_{SR}(S,R)$.

Every recursive dependency removes a point from one of the two sets, so the tables
can be filled in increasing order of $|S'|+|R'|$.
There are at most
$\sum_{S'\subseteq S}\sum_{R'\subseteq R}|S'| \cdot |R'|=\mathcal{O}(nm2^{n+m})$ states,
and the auxiliary tables are no larger.
Each non-boundary entry is evaluated in constant time from three already computed
auxiliary values and $d(s_i,r_\ell)$, and computing all auxiliary values by
scanning the endpoint pairs takes $\mathcal{O}(nm2^{n+m})$ time in total.
The boundary values, the lower sets and their maximal elements are precomputed
within the same bound, which also holds for the space.
The full proof is given in Appendix~\ref{app:proofAlgo2}.
\end{proof}
\subsection{Continuous Fréchet distance of partially ordered point sets}
\label{sec:Alg3}

\begin{figure}[t]
    \centering
    \includegraphics[width=1\textwidth]{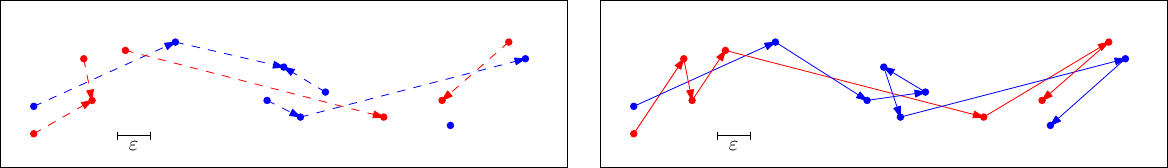}
    \caption{An instance of the continuous problem (left). Point set $S$ is red and $R$ is blue. Dashed arrows represent partial order relations. A feasible pair of curves realizing $\delta_F(S, R) \leq \varepsilon$ (right).}
    \label{fig:example_cont}
\end{figure}

We now consider the continuous Fréchet distance between two partially ordered
point sets $S=\{s_1,\ldots,s_n\}$ and $R=\{r_1,\ldots,r_m\}$.
Two things change compared to the previous subsection.
First, we may traverse a full edge (or even multiple edges) of one curve while only traversing part of an edge of the other curve. The cost of a traversal is no longer determined by the distances between matched vertices, but between all pairs of points on the curves that are visited at the same time in both traversals.
We therefore fix a threshold $\varepsilon>0$ and decide whether
$\delta_F(S,R)\le\varepsilon$, combining the subset dynamic programming algorithm with the
free-space reachability framework of Alt and Godau~\cite{AltGodau1995}. The exact
value then follows with logarithmic overhead (Theorem~\ref{thm:Algo3exact}).
Second, a state has to describe how far along its last edge a traversal currently
is, so we use three families of states and store, instead of a distance, the
smallest parameter along this edge at which the corresponding position can be
reached.
Figure~\ref{fig:example_cont} shows an example.

For $x,y\in\mathbb R^2$ we write $\overline{xy}$ for the segment oriented from
$x$ to $y$, parametrized by $\overline{xy}(t)=(1-t)x+ty$ for $t\in[0,1]$.
For an oriented segment $e$ and a point $p$, the parameters $t$ with
$d(e(t),p)\le\varepsilon$ form a single, possibly empty, closed interval, which we
describe by its endpoints $\leftI(e,p)$ and $\rightI(e,p)$. If it is empty we set
$\leftI(e,p)=\infty$ and $\rightI(e,p)=-\infty$.
In particular, all of $e$ is within distance $\varepsilon$ of $p$ if and only if
both endpoints of $e$ are.
For a finite point set $X$ we write $d_{\max}(p,X)=\max_{x\in X}d(p,x)$. 

Let $S'\in\mathcal J(S)$ and $R'\in\mathcal J(R)$ be non-empty lower sets and let
$\pi$ and $\rho$ be linear extensions of them.
A \emph{position} of the induced curves $c_\pi(S')$ and $c_\rho(R')$ is a pair of
points, one on each curve. 
It is \emph{reachable} if both curves can be traversed
continuously and monotonically from their start points to this position such that
the distance between the two moving points never exceeds $\varepsilon$.
By the definition of the Fréchet distance, $\delta_F(S,R)\le\varepsilon$ holds if
and only if for some linear extensions $\pi$ of $\prec_S$ and $\rho$ of $\prec_R$
the position consisting of the two last vertices is reachable.

As in the previous subsection, the visited points form non-empty lower sets and
the last visited point of such a set is maximal in it.
With $s_i\in\operatorname{Max}(S')$ and $r_\ell\in\operatorname{Max}(R')$
throughout, we use three families of states:
(1)~$(s_i,S',r_\ell,R')$, where both traversals are at a vertex,
(2)~$(s_i,S',\overline{r_hr_\ell},R')$ with $|R'|>1$ and
$r_h\in\operatorname{Max}(R'\setminus\{r_\ell\})$, where the traversal on $R'$ is
on its last edge while the one on $S'$ is at the vertex $s_i$,
and (3)~$(\overline{s_js_i},S',r_\ell,R')$ with $|S'|>1$ and
$s_j\in\operatorname{Max}(S'\setminus\{s_i\})$, the symmetric situation.
A pair $(\pi,\rho)$ of linear extensions of $S'$ and $R'$ is \emph{compatible}
with a state if it ends in $s_i$ and $r_\ell$ and, in the cases~(2) and~(3),
visits $r_h$ resp.\ $s_j$ immediately before.
Every state has a compatible pair, since removing a maximal element from a lower
set leaves a lower set, every maximal element of a lower set is the last element
of some linear extension of it~\cite{Stanley2011}, and appending it again yields
a linear extension.

The table entries store the following values, where the infimum ranges over all $t\in[0,1]$ and all pairs $(\pi,\rho)$ of linear extensions compatible with the state. We define $\inf\emptyset:=\infty$.
Reachability refers to the curves $c_\pi(S')$ and $c_\rho(R')$ and to the position given on the left
\begin{align*}
\continuousPO_1(s_i,S',r_\ell,R')
    &=0\quad\text{if }(s_i,r_\ell)\text{ is reachable, and }\infty
    \text{ otherwise},\\
\continuousPO_2(s_i,S',\overline{r_hr_\ell},R')
    &=\inf\bigl\{t\ \bigm|\ (s_i,\overline{r_hr_\ell}(t))\text{ is reachable}\bigr\},\\
\continuousPO_3(\overline{s_js_i},S',r_\ell,R')
    &=\inf\bigl\{t\ \bigm|\ (\overline{s_js_i}(t),r_\ell)\text{ is reachable}\bigr\}.
\end{align*}
Keeping only the smallest reachable parameter suffices, since a traversal can
always advance further along its current edge while the other one rests, and
everything that can be completed from some parameter can also be completed from a
smaller one. 
No state with both traversals inside an edge is needed, since one of them is at a
vertex whenever they move on to a new pair of edges.

Let $|S'|>1$ and $|R'|>1$.
For $s_j\in\operatorname{Max}(S'\setminus\{s_i\})$ and
$r_h\in\operatorname{Max}(R'\setminus\{r_\ell\})$ we abbreviate the three possible
predecessor values
\begin{align*}
D_1(s_j,r_h)&=\continuousPO_1(s_j,S'\setminus\{s_i\},r_h,R'\setminus\{r_\ell\}),\\
D_2(s_j,r_h)&=\continuousPO_2(s_j,S'\setminus\{s_i\},\overline{r_hr_\ell},R'),\\
D_3(s_j,r_h)&=\continuousPO_3(\overline{s_js_i},S',r_h,R'\setminus\{r_\ell\}).
\end{align*}
They correspond to the three ways in which the traversals can have arrived at the
current position while the two last edges $\overline{s_js_i}$ and
$\overline{r_hr_\ell}$ are traversed: both have left their vertices $s_j$ and
$r_h$ at the same moment, or the traversal of $R$ was already on
$\overline{r_hr_\ell}$ while the one of $S$ advanced from $s_j$ to $s_i$, or
symmetrically.
Accordingly, a term that lets the traversal of $S$ advance replaces $S'$ by
$S'\setminus\{s_i\}$ and keeps $R'$, and vice versa.
The auxiliary minima are
\begin{align*}
M_1(s_i,S',r_\ell,R')&=\min_{s_j,r_h}\ \min\{D_1,D_2,D_3\},\\
M_2(s_i,S',\overline{r_hr_\ell},R')&=\min_{s_j}\ \min\{D_1,D_2,\zeta(D_3)\},\\
M_3(\overline{s_js_i},S',r_\ell,R')&=\min_{r_h}\ \min\{D_1,\zeta(D_2),D_3\},
\end{align*}
where $s_j$ and $r_h$ range as above, except that $r_h$ is given by the state in
$M_2$ and $s_j$ is given by the state in $M_3$.
The value carried over is the parameter along the edge on which the traversal in
question was already moving. 
In the other two terms this traversal enters its
edge at its start point, which is expressed by $\zeta(x)=0$ for $x\neq\infty$ and
$\zeta(\infty)=\infty$.
With these, and abbreviating $M_1$, $M_2$ and $M_3$ for the auxiliary value of
the state on the left, the recurrences are
\begin{align*}
\continuousPO_1(s_i,S',r_\ell,R')
    &=
    \begin{cases}
    0,&\text{if } d(s_i,r_\ell)\le\varepsilon \text{ and } M_1\neq\infty,\\
    \infty,&\text{otherwise,}
    \end{cases}
    \\[1ex]
\continuousPO_2(s_i,S',\overline{r_hr_\ell},R')
    &=
    \begin{cases}
    \max\bigl\{M_2,\ \leftI(\overline{r_hr_\ell},s_i)\bigr\},
    &\text{if } M_2\le \rightI(\overline{r_hr_\ell},s_i),\\
    \infty,&\text{otherwise,}
    \end{cases}
    \\[1ex]
\continuousPO_3(\overline{s_js_i},S',r_\ell,R')
    &=
    \begin{cases}
    \max\bigl\{M_3,\ \leftI(\overline{s_js_i},r_\ell)\bigr\},
    &\text{if } M_3\le \rightI(\overline{s_js_i},r_\ell),\\
    \infty,&\text{otherwise.}
    \end{cases}
\end{align*}
The carried parameter is clamped into the interval of parameters within distance
$\varepsilon$ of the opposite vertex: below $\leftI$ the traversal can advance to
$\leftI$, and above $\rightI$ the position cannot be reached at all.
For a state of type~(1) only the finiteness of $M_1$ matters, since both
traversals can move to the ends $s_i$ and $r_\ell$ of their current edges once
these are within distance $\varepsilon$ of each other.

In a boundary case, that is for $|S'|=1$ or $|R'|=1$, one of the two curves is a
single point, so the traversal of the other curve has to keep everything
traversed so far within distance $\varepsilon$ of this point, and by the above it
suffices to check the vertices.
The order on the other side is then irrelevant and the traversal can stay at the
start of its current edge, so all finite boundary values are $0$:
\begin{align*}
\continuousPO_1(s_i,\{s_i\},r_\ell,R')&=0\quad\text{if } d_{\max}(s_i,R')\le\varepsilon,\\
\continuousPO_1(s_i,S',r_\ell,\{r_\ell\})&=0\quad\text{if } d_{\max}(r_\ell,S')\le\varepsilon,\\
\continuousPO_2(s_i,\{s_i\},\overline{r_hr_\ell},R')&=0\quad\text{if } d_{\max}(s_i,R'\setminus\{r_\ell\})\le\varepsilon,\\
\continuousPO_3(\overline{s_js_i},S',r_\ell,\{r_\ell\})&=0\quad\text{if } d_{\max}(r_\ell,S'\setminus\{s_i\})\le\varepsilon,
\end{align*}
and $\infty$ otherwise.
For $|S'|=|R'|=1$ the first two lines agree and state that the traversals can
start at all.

A dynamic programming algorithm fills the tables in increasing order of $|S'|+|R'|$, starting
with the boundary cases. Every recurrence formula requires only strictly smaller values of this quantity.
The algorithm then answers \emph{yes} if and only if
$\continuousPO_1(s_i,S,r_\ell,R)=0$ for some $s_i\in\operatorname{Max}(S)$ and
some $r_\ell\in\operatorname{Max}(R)$.

\begin{theorem}
\label{thm:Algo3}
    Let $S$ and $R$ be partially ordered point sets of sizes $n$ and $m$ and let
    $\varepsilon>0$.
    Then $\continuousPO_1(s_i,S,r_\ell,R)=0$ for some
    $s_i\in\operatorname{Max}(S)$ and $r_\ell\in\operatorname{Max}(R)$ if and only
    if $\delta_F(S,R)\le\varepsilon$.
    All table entries can be computed in $\mathcal{O}(n^2m^22^{n+m})$ time and
    $\mathcal{O}(nm(n+m)2^{n+m})$ space.
\end{theorem}

\begin{proof}[sketch]
Correctness is shown by induction on $|S'|+|R'|$ in the free space diagram of a
fixed pair of induced curves, in which the three families of states are the top right corner, the right edge and the top edge of the cell spanned by the two last
edges.
In the base cases one of the curves is a single point, so the order on the other
side is irrelevant.
Otherwise, a monotone path reaching the current position enters this cell at its bottom left corner, through
its left edge, or through its bottom edge, which correspond to $D_1$, $D_2$ and $D_3$. 
Removing a maximal element from a lower set
leaves a lower set, so all three are again states.
Since the free space inside a cell is convex, every reachable entry point can be
connected to the current position by a monotone path as soon as that position is
free, so keeping the smallest parameter and clamping it is correct.
Conversely, each term yields a feasible pair of induced curves, since appending a
maximal element to a linear extension of the remaining lower set again gives a
linear extension.
Minimizing over the last vertices gives the claim.

There are $\mathcal{O}(nm2^{n+m})$ states of type~(1), and a state of type~(2)
resp.~(3) is such a state together with the point $r_h$ resp.~$s_j$, so the
tables have $\mathcal{O}(nm(n+m)2^{n+m})$ entries.
They are evaluated by minimizing over $\mathcal{O}(nm)$, $\mathcal{O}(n)$ and
$\mathcal{O}(m)$ predecessors, respectively, which sums to
$\mathcal{O}(n^2m^22^{n+m})$.
Precomputing the lower sets with their maximal
elements, the values of $\leftI$ and $\rightI$, and the values $d_{\max}$ is
dominated by this.
The full proof is given in Appendix~\ref{app:proofAlgo3}.
\end{proof}

The exact value is found by searching a set of critical values that, crucially,
does not depend on the chosen linear extensions, since the edges of any induced
curve are spanned by two points of the same input set.

\begin{theorem}
\label{thm:Algo3exact}
    Given partially ordered point sets $S$ and $R$ of sizes $n$ and $m$, the
    value $\delta_F(S,R)$ can be computed in
    $\mathcal{O}(n^2m^22^{n+m}\log(nm))$ time.
\end{theorem}

\begin{proof}[sketch]
For two fixed curves the Fréchet distance is one of three types of critical
values~\cite{AltGodau1995}: the distance of the two start points or of the two
endpoints, the distance of a vertex of one curve to an edge of the other, or the
common distance of two vertices of one curve to the intersection of an edge of
the other curve with their perpendicular bisector.
Each of them is determined by at most two points of $S$ and two points of $R$, so
the set $\mathcal C$ of all values of these three types is the same for all pairs
of induced curves, has size $\mathcal{O}(n^2m^2)$ and can be computed in
$\mathcal{O}(n^2m^2)$ time.
As $\delta_F(S,R)$ is the minimum of $\delta_F(c_\pi(S),c_\rho(R))$ over the
finitely many pairs $(\pi,\rho)$, it lies in $\mathcal C$, and since the predicate
$\delta_F(S,R)\le\varepsilon$ is monotone in $\varepsilon$, a binary search over
the sorted set $\mathcal C$ finds it with $\mathcal{O}(\log(nm))$ calls of the
decision procedure of Theorem~\ref{thm:Algo3}.
Sorting $\mathcal C$ is dominated by a single such call.
The full proof is given in Appendix~\ref{app:proofAlgo3exact}.
\end{proof}

\paragraph{Practical aspects.}
Several simple measures save time and space in practice without improving the
worst case. Every recursive dependency removes one or two points, so, as in
Theorem~\ref{thm:Algo1}, only the entries for three consecutive values of
$|S'|+|R'|$ have to be kept if only the value is needed. For the decision
problems, the tables can be filled forward from the start states, so that only
states with finite value are generated, which can be far fewer than all states.

Finally, $\delta_F(S,R)\le\delta_{dF}(S,R)$, since a discrete coupling of two curves
yields a traversal with the same cost; hence the binary search of
Theorem~\ref{thm:Algo3exact} can be restricted to the values of $\mathcal C$ that
are at most $\delta_{dF}(S,R)$, which Theorem~\ref{thm:Algo2} computes faster than a
single call of the decision procedure.

Lower bounds for the exponential factor in the two theorems above used to rely
on ETH~\cite{ImpagliazzoPaturi2001}, but deterministic $2^{o(v)}$-time
algorithms for 3-SAT on $v$ variables have been
announced~\cite{Ope26,Ant26}. Their running time $2^{v^{1-o(1)}}$
is still consistent with the \emph{Satisfiability Hypothesis} (SH) of Stearns and
Hunt~\cite{StearnsHunt1990,MIT26}: for no constant $\gamma>0$ can CNF-SAT of length $L$ be decided in time
$\mathcal O(2^{L^{1-\gamma}})$. Hence we now formulate a lower bound based on SH. 

\begin{proposition}\label{prop:eth}
A 3-SAT formula with $c$ clauses can be transformed in polynomial time into point
sets $S,R\subset\mathbb Q^2$ with $|S|+|R|=\mathcal O(c)$ and a value
$\varepsilon>0$ such that the formula is satisfiable if and only if
$\delta_F(S,R)\le\varepsilon$.
\end{proposition}

\begin{corollary}\label{cor:sh}
Assuming SH, for no constant $\gamma>0$ is there a deterministic algorithm that
decides for point sets $S,R\subset\mathbb Q^2$ with $|S|=n$, $|R|=m$ and
$\varepsilon\in\mathbb Q_{>0}$, given in binary with $B$ bits in total, whether
$\delta_F(S,R)\le\varepsilon$ in time
$2^{\mathcal O((n+m)^{1-\gamma})}\operatorname{poly}(B)$. The same holds for
partially ordered point sets, so the exponent $n+m$ in Theorems~\ref{thm:Algo3}
and~\ref{thm:Algo3exact} cannot be improved to $(n+m)^{1-\gamma}$.
\end{corollary}

\begin{proof}[sketch]
The reduction of Buchin and Kilgus~\cite{BUCHIN2022} from $(3,B2)$-SAT, in which
every literal occurs exactly twice, produces $n+m=\frac{34}{3}V$ points for a
formula with $V$ variables. Via NAE-3SAT and the enforcers of
Berman, Karpinski and Scott~\cite{BermanKarpinskiScott2004}, a 3-SAT formula with
$c$ clauses becomes a $(3,B2)$-SAT formula with $V=30c$ variables, so
$n+m=340c$. The coordinates are
rationals of $\mathcal O(\log c)$ bits. Splitting long clauses
turns a CNF formula of length $L$ into a 3-SAT formula with at most $L$ clauses,
so an algorithm as excluded in Corollary~\ref{cor:sh} would decide CNF-SAT in
time $2^{\mathcal O(L^{1-\gamma})}$, contradicting SH.
For partially ordered point sets we take empty orders. Details are in
Appendix~\ref{app:eth}.
\end{proof}
\section{Conclusion}

We generalized the discrete and continuous Fréchet distance to partially ordered point sets by optimizing over all linear extensions of the input orders, and gave exact subset-based dynamic-programming algorithms for the resulting problems.
We also showed that, under the Satisfiability Hypothesis,
$\delta_F(S,R)\le\varepsilon$ cannot be decided in time
$2^{\mathcal O((n+m)^{1-\gamma})}$ for any constant $\gamma>0$.

Several questions remain open.
Does a lower bound of this form also hold for the discrete variant?
Is time $2^{o(n+m)}$ possible, as claimed for 3-SAT~\cite{Ope26,Ant26}?
The Frechet distance of partially ordered points interpolates between two polynomially solvable problems. If both orders are total, the distance is computed in $\mathcal O(nm)$ time~\cite{EiterMannila1994}, and if both orders are empty, it equals the Hausdorff distance and is computed in $\mathcal O((n+m)\log(nm))$ time~\cite{BUCHIN2022}, while already the intermediate case of one total and one empty order is NP-hard~\cite{WylieZhu2013}.
It would be interesting to understand where exactly this transition happens: which pairs of classes of partial orders admit polynomial-time algorithms?
Moreover, if both partial orders have width at most $w$, then by Dilworth's theorem each is covered by $w$ chains, and a lower set is determined by a prefix of each of them, so $|\mathcal J(X)|\le(|X|/w+1)^w$ and our algorithms run in time $(n+m)^{\mathcal O(w)}$, that is, they are XP in the width.
Are they fixed-parameter tractable in the width, or is the problem W[1]-hard for this parameter?
Combining our model with spatial uncertainty~\cite{AhnKSSV2012,BuchinFLPRR2023} is another direction.
Finally, it would be interesting to evaluate the exact algorithms experimentally,
exploring only reachable states, and to develop heuristics for larger instances.

\bibliographystyle{alpha}
\bibliography{Inputs/Frechet_distances_of_Point_Sets}

\appendix
\renewcommand{\theHsection}{A\arabic{section}}

\section{Omitted Proofs}
\label{app:proofs}

\subsection{Proof of Theorem~\ref{thm:Algo1}}
\label{app:proofAlgo1}
\begin{proof}
\textbf{Correctness.}
For $s_i \in S'$, let $\Delta(s_i, S', k)$ denote the minimum discrete Fréchet distance between $P_k$ and any curve induced by $S'$ whose last vertex is $s_i$.
We show by induction on $|S'|+k$ that $\dusm(s_i,S',k)=\Delta(s_i,S',k)$ for all non-empty $S'\subseteq S$, all $s_i\in S'$, and all
$k\in\{1,\ldots,m\}$.
After the values $\dusm(s_i,S',k)$ for a fixed pair $(S',k)$ are computed, $M(S',k)$ is their minimum over all possible endpoints $s_i\in S'$.
Hence $M(S',k)$ is correct whenever the corresponding $\dusm$ values are
correct.

If $|S'|=1$, then $S'=\{s_i\}$ and the only induced curve consists of the single
point $s_i$.
Thus every point of $P_k$ must be matched to $s_i$, which gives
$\Delta(s_i,\{s_i\},k)=\max_{1\le j\le k} d(s_i,p_j)=\dusm(s_i,\{s_i\},k)$ for all $s_i \in S$ and $k\leq m$.
If $k=1$, then $P_k$ consists only of $p_1$.
Hence every point of the induced curve on $S'$ must be matched to $p_1$,
independently of the order chosen on $S'$.
Therefore $\Delta(s_i,S',1)=\max_{s_j\in S'} d(s_j,p_1)=\dusm(s_i,S',1)$ holds for all non-empty $S' \subseteq S$ and $s_i \in S'$.
This proves the claim in the base cases, where $|S'|=1$ or $k=1$.

Now let $|S'|>1$ and $k>1$, and assume the claim holds for all smaller values of
$|S'|+k$.
Consider an optimal coupling between $P_k$ and an induced curve on $S'$ ending
in $s_i$.
Its last pair must be $(s_i,p_k)$.
After deleting this last pair, the remaining coupling ends in a pair obtained by
decreasing at least one of the two indices.
Thus the predecessor of $(s_i,p_k)$ is of exactly one of three types, see Figure~\ref{fig:predecessor-types}.

\begin{figure}[b]
  \centering
  \includegraphics[width=\textwidth]{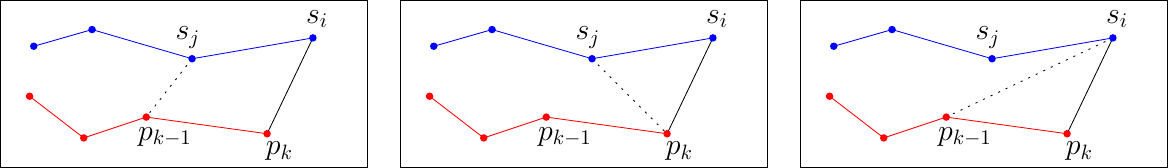}
  \caption{The three possible predecessors of the last pair $(s_i,p_k)$ in a
    coupling between $P_k$ (red) and a curve induced by $S'$ that ends in $s_i$
    (blue). The last pair is drawn solid and its predecessor dotted. Left:
    $(s_j,p_{k-1})$, covered by $M(S'\setminus\{s_i\},k-1)$. Middle: $(s_j,p_k)$,
    covered by $M(S'\setminus\{s_i\},k)$. Right: $(s_i,p_{k-1})$, covered by
    $\delta_{\mathrm{USM}}(s_i,S',k-1)$. The three cases in the proof of
    Theorem~2 are the same, with $p_{k-1},p_k$ replaced by $r_h,r_\ell$.}
  \label{fig:predecessor-types}
\end{figure}

First, both indices may decrease.
Then the predecessor is $(s_j,p_{k-1})$ for some
$s_j \in S' \setminus \{s_i\}$.
The remaining coupling is a coupling between $P_{k-1}$ and a curve induced by
$S' \setminus \{s_i\}$ ending in $s_j$.
By the induction hypothesis and the definition of $M$, the best possible value
of such a predecessor coupling is $M(S'\setminus\{s_i\},k-1)$.

Second, only the index in the induced curve may decrease.
Then the predecessor is $(s_j,p_k)$ for some
$s_j \in S' \setminus \{s_i\}$.
The remaining coupling is a coupling between $P_k$ and a curve induced by
$S' \setminus \{s_i\}$ ending in $s_j$.
Again by the induction hypothesis and the definition of $M$, the best possible
predecessor value in this case is $M(S'\setminus\{s_i\},k)$.

Third, only the index in $P_k$ may decrease.
Then the predecessor is $(s_i,p_{k-1})$.
The remaining coupling is a coupling between $P_{k-1}$ and a curve induced by
$S'$ that still ends in $s_i$.
By the induction hypothesis, the best possible predecessor value in this case is
$\dusm(s_i,S',k-1)$.

For any fixed predecessor coupling of value $L$, appending the final pair
$(s_i,p_k)$ gives a coupling of value $\max\{L,d(s_i,p_k)\}$.
The three predecessor cases form a partition of all possible couplings whose
last pair is $(s_i,p_k)$.
Therefore the optimum value is obtained by taking the best predecessor among
the three cases and then accounting for the distance of the final pair.
This is exactly the recurrence defining $\dusm(s_i,S',k)$.
Hence $\dusm(s_i,S',k)=\Delta(s_i,S',k)$, completing the induction.

Finally, the discrete Fréchet distance between $S$ and $P$ is obtained by
choosing the endpoint of the induced curve on $S$ optimally.
Thus, we have $\delta_{dF}(S,P)=\min_{s_i\in S}\Delta(s_i,S,m)=M(S,m)$.

\textbf{Complexity.}
Every recursive dependency of $\dusm(s_i,S',k)$ lies in layer $k-1$, namely
$M(S'\setminus\{s_i\},k-1)$ and $\dusm(s_i,S',k-1)$, or in layer $k$ at a
strictly smaller set, namely $M(S'\setminus\{s_i\},k)$.
The table of the dynamic programming algorithm can therefore be filled in increasing order of $k$ and, for
equal $k$, in increasing order of $|S'|$.
The boundary values for singleton subsets can be computed using prefix maxima
of the distances $d(s_i,p_k)$.
The boundary values for $k=1$ can be initialized by precomputing
$\max_{s_j\in S'}d(s_j,p_1)$ for all subsets $S'\subseteq S$.
These initializations fit within the final asymptotic running time.

The number of entries $\dusm(s_i,S',k)$ is bounded by the number of triples
$(s_i,S',k)$ with $s_i\in S'\subseteq S$ and $k\le m$.
For each fixed $k$, this number is $\sum_{S'\subseteq S}|S'|=\mathcal{O}(n2^n)$,
so over all $m$ values of $k$ there are $\mathcal{O}(nm2^n)$ such entries.
There are additionally $\mathcal{O}(m2^n)$ entries of the auxiliary table $M$.

For a fixed entry $\dusm(s_i,S',k)$ with $|S'|>1$ and $k>1$, the recurrence uses
only three already computed values, namely $M(S'\setminus\{s_i\},k-1)$,
$M(S'\setminus\{s_i\},k)$, and $\dusm(s_i,S',k-1)$.
Thus each recurrence evaluation takes constant time, assuming distances have
been precomputed or can be evaluated in constant time.
Computing all values $M(S',k)$ by scanning the endpoints $s_i\in S'$ takes total
time $\sum_{k=1}^m \sum_{S'\subseteq S}|S'|=\mathcal{O}(nm2^n)$.
Hence the total running time is $\mathcal{O}(nm2^n)$.

The table $\dusm$ uses $\mathcal{O}(nm2^n)$ space and the auxiliary table $M$
uses $\mathcal{O}(m2^n)$ additional space, so storing all layers takes
$\mathcal{O}(nm2^n)$ space.
This can be reduced to $\mathcal{O}(n2^n)$.
By the fill order above, the entries of layer $k$ depend only on layers $k$ and
$k-1$, so it suffices to keep these two layers and to discard layer $k-1$ once
layer $k$ is complete.
A single layer consists of $\mathcal{O}(n2^n)$ entries of $\dusm$ and
$\mathcal{O}(2^n)$ entries of $M$, and the boundary values of layer $k$ are
obtained from those of layer $k-1$ within the window, since
$\dusm(s_i,\{s_i\},k)=\max\{\dusm(s_i,\{s_i\},k-1),d(s_i,p_k)\}$ and the values
for $k=1$ belong to the first layer.
The distances $d(s_i,p_k)$ are evaluated on demand on the real RAM, so no
distance table is needed.
Hence the working space is $\mathcal{O}(n2^n)$, excluding the
$\mathcal{O}(n+m)$ space occupied by the input.
Note that this applies to computing the value $M(S,m)$; reconstructing an
optimal ordering and an optimal coupling by backtracking requires the
predecessor choices of all layers and hence $\mathcal{O}(nm2^n)$ space. \qed
\end{proof}

\subsection{Proof of Theorem~\ref{thm:Algo2}}
\label{app:proofAlgo2}
\begin{proof}
\textbf{Correctness.}
For a state as above, let $\Delta(s_i,S',r_\ell,R')$ denote the minimum discrete
Fréchet distance between any curve induced by $S'$ ending in $s_i$ and any curve
induced by $R'$ ending in $r_\ell$.
Since $S'$ and $R'$ are lower sets and the two endpoints are maximal in their
respective sets, such curves exist, so this value is well defined and finite.
We show by induction on $|S'|+|R'|$ that
${\ddiscretePO}(s_i,S',r_\ell,R')=\Delta(s_i,S',r_\ell,R')$ for all states.
Correctness of $M_{SR}$, $M_S$, and $M_R$ then follows, since they are minima of
correct values over the remaining endpoint choices.

In the base cases the order on one of the two sets is irrelevant.
If $|S'|=1$, then $S'=\{s_i\}$ and the only curve induced by $S'$ consists of the
single point $s_i$.
Every coupling therefore matches all points of $R'$ to $s_i$, independently of
the linear extension chosen on $R'$.
This gives $\Delta(s_i,\{s_i\},r_\ell,R')=\max_{r_h\in R'} d(s_i,r_h)= {\ddiscretePO}(s_i,\{s_i\},r_\ell,R')$.
The case $|R'|=1$ is symmetric and gives
$\Delta(s_i,S',r_\ell,\{r_\ell\})=\max_{s_j\in S'} d(s_j,r_\ell)={\ddiscretePO}(s_i,S',r_\ell,\{r_\ell\})$.
This proves the claim in the base cases, where $|S'|=1$ or $|R'|=1$.

Now let $|S'|>1$ and $|R'|>1$, and assume the claim holds for all smaller values
of $|S'|+|R'|$.
Consider an optimal coupling between a curve induced by $S'$ ending in $s_i$ and
a curve induced by $R'$ ending in $r_\ell$.
Its last pair must be $(s_i,r_\ell)$.
After deleting this last pair, the remaining coupling ends in a pair obtained by
decreasing at least one of the two indices.
Thus the predecessor of $(s_i,r_\ell)$ is of exactly one of three types.

First, both indices may decrease.
Then the predecessor is $(s_j,r_h)$ for some $s_j\in S'\setminus\{s_i\}$ and
$r_h\in R'\setminus\{r_\ell\}$.
The remaining coupling is a coupling between a curve induced by
$S'\setminus\{s_i\}$ ending in $s_j$ and a curve induced by
$R'\setminus\{r_\ell\}$ ending in $r_h$.
Both sets are lower sets, and $s_j$ and $r_h$ are maximal in them, so the
corresponding entries are states.
By the induction hypothesis and the definition of $M_{SR}$, the best possible
value of such a predecessor coupling is
$M_{SR}(S'\setminus\{s_i\},R'\setminus\{r_\ell\})$.

Second, only the index in the induced curve on $S'$ may decrease.
Then the predecessor is $(s_j,r_\ell)$ for some $s_j\in S'\setminus\{s_i\}$.
The remaining coupling is a coupling between a curve induced by
$S'\setminus\{s_i\}$ ending in $s_j$ and a curve induced by $R'$ ending in
$r_\ell$.
By the induction hypothesis and the definition of $M_S$, the best possible
predecessor value in this case is $M_S(S'\setminus\{s_i\},r_\ell,R')$.

Third, only the index in the induced curve on $R'$ may decrease.
Then the predecessor is $(s_i,r_h)$ for some $r_h\in R'\setminus\{r_\ell\}$.
The remaining coupling is a coupling between a curve induced by $S'$ ending in
$s_i$ and a curve induced by $R'\setminus\{r_\ell\}$ ending in $r_h$.
By the induction hypothesis and the definition of $M_R$, the best possible
predecessor value in this case is $M_R(s_i,S',R'\setminus\{r_\ell\})$.

These three cases partition all couplings whose last pair is $(s_i,r_\ell)$.
Conversely, every coupling counted by one of the three terms can be extended by
the pair $(s_i,r_\ell)$.
On the side of $S$, appending $s_i$ to a linear extension of $S'\setminus\{s_i\}$
gives a linear extension of $S'$, because $s_i$ is maximal in $S'$.
The same argument applies on the side of $R$.
Thus the three terms range exactly over the feasible predecessor couplings.
Since appending $(s_i,r_\ell)$ to a predecessor coupling of value $L$ yields a
coupling of value $\max\{L,d(s_i,r_\ell)\}$, taking the best predecessor among
the three cases and then the maximum with $d(s_i,r_\ell)$ is exactly the
recurrence.
Hence ${\ddiscretePO}(s_i,S',r_\ell,R')=\Delta(s_i,S',r_\ell,R')$, which
completes the induction.
Finally, an optimal pair of induced curves on $S$ and $R$ has some last vertices
$s_i\in\operatorname{Max}(S)$ and $r_\ell\in\operatorname{Max}(R)$.
Minimizing over these endpoints gives $\delta_{dF}(S,R)=M_{SR}(S,R)$.

\textbf{Complexity.}
Every recursive dependency removes a point from one of the two sets and
therefore decreases $|S'|+|R'|$, so the table can be filled in increasing order
of this quantity.
The number of entries of ${\ddiscretePO}$ is the number of tuples
$(s_i,S',r_\ell,R')$ with $S'\in\mathcal J(S)$, $R'\in\mathcal J(R)$,
$s_i\in\operatorname{Max}(S')$ and $r_\ell\in\operatorname{Max}(R')$.
This is at most
$\sum_{S'\subseteq S}\sum_{R'\subseteq R}|S'||R'|=\mathcal{O}(nm2^{n+m})$.
The auxiliary tables have at most the same number of entries.

Each non-boundary entry is evaluated in constant time from three already
computed auxiliary values and the distance $d(s_i,r_\ell)$, assuming distances
are available in constant time.
For a fixed pair $(S',R')$, all values of $M_{SR}$, $M_S$, and $M_R$ are
computed by scanning the endpoint pairs in
$\operatorname{Max}(S')\times\operatorname{Max}(R')$, which over all pairs
$(S',R')$ takes $\mathcal{O}(nm2^{n+m})$ time in total.
The boundary values are obtained from subset maxima of $d(s_i,\cdot)$ for each
fixed $s_i\in S$ and of $d(\cdot,r_\ell)$ for each fixed $r_\ell\in R$, which
takes $\mathcal{O}(n2^m+m2^n)$ time.
The family of lower sets and the sets of maximal elements are precomputed by
a dynamic programming algorithm over bit masks in $\mathcal{O}(n2^n+m2^m)$ bit-mask
operations.
All these bounds are dominated by $\mathcal{O}(nm2^{n+m})$, which is therefore
the total running time.

The table ${\ddiscretePO}$ uses $\mathcal{O}(nm2^{n+m})$ space.
The auxiliary tables use $\mathcal{O}((n+m)2^{n+m})$ additional space, which is
dominated by $\mathcal{O}(nm2^{n+m})$ for non-empty input posets.
Thus the total space usage is $\mathcal{O}(nm2^{n+m})$. \qed
\end{proof}

\subsection{Proof of Theorem~\ref{thm:Algo3}}
\label{app:proofAlgo3}

For the proof we use the free-space diagram of Alt and
Godau~\cite{AltGodau1995}.
Let $S'\in\mathcal J(S)$ and $R'\in\mathcal J(R)$ be non-empty lower sets and let
$\pi$ and $\rho$ be linear extensions of $S'$ and $R'$.
The induced curves are the maps $c_\pi(S')\colon[1,|S'|]\to\mathbb R^2$ and
$c_\rho(R')\colon[1,|R'|]\to\mathbb R^2$, where a one-point curve is the constant
map on the degenerate interval $[1,1]$.
Their \emph{free-space} is
$$
F_\varepsilon(\pi,\rho)
=
\bigl\{(u,v)\in[1,|S'|]\times[1,|R'|]
\ \bigm|\ d\bigl(c_\pi(S')(u),c_\rho(R')(v)\bigr)\le\varepsilon\bigr\}.
$$
A point $(u,v)\in F_\varepsilon(\pi,\rho)$ is \emph{reachable} if there is a
continuous path inside $F_\varepsilon(\pi,\rho)$ from $(1,1)$ to $(u,v)$ that is
non-decreasing in both coordinates. 
We call such a path \emph{monotone} and
write $\mathcal R(\pi,\rho)$ for the set of all reachable points.
A position of the two induced curves in the sense of Section~\ref{sec:Alg3} is
reachable if and only if the corresponding parameter pair lies in
$\mathcal R(\pi,\rho)$, since the traversals of the two curves are exactly the
two coordinates of a monotone path.
For a state we write $\mathcal P$ of the state, for example $\mathcal P(s_i,S',\overline{r_hr_\ell},R')$, for the set of pairs $(\pi,\rho)$ of linear extensions that are compatible with it in the sense of Section~\ref{sec:Alg3}.
In these terms the table entries store the values
\begin{align*}
\continuousPO_1(s_i,S',r_\ell,R')
    &=
    \begin{cases}
    0,&\text{if }(|S'|,|R'|)\in\mathcal R(\pi,\rho),\\
    \infty,&\text{otherwise,}
    \end{cases}
    \\[1ex]
\continuousPO_2(s_i,S',\overline{r_hr_\ell},R')
    &=
    \inf\bigl\{t\ \bigm|\ (|S'|,\,|R'|-1+t)\in\mathcal R(\pi,\rho)\bigr\},
    \\[1ex]
\continuousPO_3(\overline{s_js_i},S',r_\ell,R')
    &=
    \inf\bigl\{t\ \bigm|\ (|S'|-1+t,\,|R'|)\in\mathcal R(\pi,\rho)\bigr\},
\end{align*}
where again $t$ ranges over $[0,1]$, the pair $(\pi,\rho)$ ranges over the set
$\mathcal P$ of the state on the left, and $\inf\emptyset=\infty$.

The free-space decomposes into the cells $[i,i+1]\times[j,j+1]$, one for each
pair consisting of an edge of $c_\pi(S')$ and an edge of $c_\rho(R')$.
We use two standard properties of the free-space~\cite{AltGodau1995}.
\begin{description}
    \item[(F1)] For a segment $e$ and a point $p$, the set
    $\{t\in[0,1]\mid d(e(t),p)\le\varepsilon\}$ is a closed, possibly empty,
    interval with endpoints $\leftI(e,p)$ and $\rightI(e,p)$, being the preimage
    of a disk under an affine map.
    In particular the distance from $p$ to a segment attains its maximum at an
    endpoint of the segment.
    \item[(F2)] For two segments $e,f$ the set
    $\{(t,t')\in[0,1]^2\mid d(e(t),f(t'))\le\varepsilon\}$ is convex.
    Consequently, if two points of a cell of the free-space are comparable
    coordinatewise, then the straight segment between them is a monotone path
    inside the free-space.
\end{description}

We also use the following observation about prefixes.
Let $S'$ be a non-empty lower set, let $s_i\in\operatorname{Max}(S')$, let $\pi$
be a linear extension of $S'$ ending in $s_i$ and let $\pi'$ be its restriction
to $S'\setminus\{s_i\}$, which is a linear extension of the lower set
$S'\setminus\{s_i\}$.
Then for every linear extension $\rho$ of a lower set $R'$ we have
$$
F_\varepsilon(\pi',\rho)
=
F_\varepsilon(\pi,\rho)\cap\bigl([1,|S'|-1]\times[1,|R'|]\bigr),
$$
since the two curves agree on $[1,|S'|-1]$.
Moreover, a point of this strip is reachable in $F_\varepsilon(\pi',\rho)$ if and
only if it is reachable in $F_\varepsilon(\pi,\rho)$, because a monotone path
ending in the strip never leaves it.
Conversely, appending $s_i$ to a linear extension of $S'\setminus\{s_i\}$ yields
a linear extension of $S'$ ending in $s_i$.
The analogous statements hold on the side of $R$.
We refer to this as the \emph{prefix property}.
Throughout the proof we use the abbreviations $D_1(s_j,r_h)$, $D_2(s_j,r_h)$ and
$D_3(s_j,r_h)$ of Section~\ref{sec:Alg3} for the three predecessor values of the
state under consideration.

\begin{proof}[of Theorem~\ref{thm:Algo3}]
\textbf{Correctness.}
We show by induction on $|S'|+|R'|$ that every table entry equals the value
defined above and that the infimum defining it is attained whenever it is finite.
For a state of type~(2) this amounts to the two statements
\begin{enumerate}
    \item[(i)] if $(|S'|,|R'|-1+t)$ is reachable in $F_\varepsilon(\pi,\rho)$ for
    some $(\pi,\rho)\in\mathcal P(s_i,S',\overline{r_hr_\ell},R')$ and some
    $t\in[0,1]$, then $\continuousPO_2(s_i,S',\overline{r_hr_\ell},R')\le t$, and
    \item[(ii)] if $\continuousPO_2(s_i,S',\overline{r_hr_\ell},R')=t<\infty$,
    then $(|S'|,|R'|-1+t)$ is reachable in $F_\varepsilon(\pi,\rho)$ for some
    $(\pi,\rho)\in\mathcal P(s_i,S',\overline{r_hr_\ell},R')$,
\end{enumerate}
and to the analogous statements for the other two types of states.

\emph{Base cases.}
Let $S'=\{s_i\}$, so that the parameter space of the curve on $S'$ is the single
point $1$ and every monotone path is vertical.
Hence $(1,v)$ is reachable in $F_\varepsilon(\pi,\rho)$ if and only if
$d(s_i,c_\rho(R')(v'))\le\varepsilon$ for all $v'\le v$.
For $v=|R'|$ this means that all of $c_\rho(R')$ lies within distance
$\varepsilon$ of $s_i$, which by~(F1) holds if and only if
$d_{\max}(s_i,R')\le\varepsilon$, independently of $\rho$; this is the first
boundary case of $\continuousPO_1$.
For a state of type~(2) with $S'=\{s_i\}$, the condition is monotone in $t$, so
either no parameter is reachable or the smallest reachable one is $t=0$; and
$t=0$ is reachable if and only if the part of $c_\rho(R')$ up to the vertex $r_h$
lies within distance $\varepsilon$ of $s_i$, which by~(F1) holds if and only if
$d_{\max}(s_i,R'\setminus\{r_\ell\})\le\varepsilon$.
This is the boundary case of $\continuousPO_2$, and it is again independent of
$\rho$.
The cases $R'=\{r_\ell\}$ are symmetric.

\emph{Induction step for states of type~(2).}
Let $|S'|>1$ and $|R'|>1$ and let
$C=[|S'|-1,|S'|]\times[|R'|-1,|R'|]$ be the cell of the two last edges.

For~(i), let $(\pi,\rho)\in\mathcal P(s_i,S',\overline{r_hr_\ell},R')$ and let
$\gamma$ be a monotone path in $F_\varepsilon(\pi,\rho)$ from $(1,1)$ to
$q=(|S'|,|R'|-1+t)$.
Let $s_j$ be the point preceding $s_i$ in $\pi$; then
$s_j\in\operatorname{Max}(S'\setminus\{s_i\})$, so $s_j$ is one of the points
over which $M_2$ minimizes.
Since $q\in C$ and $C$ is closed, $\tau=\min\{a\mid\gamma(a)\in C\}$ is well
defined.
If $\tau>0$, then by minimality the first coordinate of $\gamma(\tau)$ equals
$|S'|-1$ or its second coordinate equals $|R'|-1$.
In the first case write $\gamma(\tau)=(|S'|-1,|R'|-1+t_0)$ with $t_0\in[0,1]$.
The restriction of $\gamma$ to $[0,\tau]$ stays in the strip
$[1,|S'|-1]\times[1,|R'|]$, so by the prefix property $t_0$ is reachable for the
pair $(\pi|_{S'\setminus\{s_i\}},\rho)$, which lies in
$\mathcal P(s_j,S'\setminus\{s_i\},\overline{r_hr_\ell},R')$.
By the induction hypothesis $D_2\le t_0$, hence $M_2\le t_0$.
In the second case write $\gamma(\tau)=(|S'|-1+u_0,|R'|-1)$ and set $t_0=0$.
By the prefix property, $u_0$ is reachable for the pair
$(\pi,\rho|_{R'\setminus\{r_\ell\}})\in
\mathcal P(\overline{s_js_i},S',r_h,R'\setminus\{r_\ell\})$, so by the induction
hypothesis $D_3\le u_0<\infty$, and therefore $M_2\le\zeta(D_3)=0=t_0$.
If $\tau=0$, then $(1,1)\in C$, hence $|S'|=|R'|=2$,
$S'\setminus\{s_i\}=\{s_j\}$ and $R'\setminus\{r_\ell\}=\{r_h\}$; since $(1,1)$
is free we have $d(s_j,r_h)\le\varepsilon$ and thus
$D_1=0$, so again $M_2\le t_0=0$.
In all three cases $M_2\le t_0\le t$, the last inequality because $\gamma$ is
monotone on $[\tau,1]$.
Moreover $q$ is free, so by~(F1)
$\leftI(\overline{r_hr_\ell},s_i)\le t\le\rightI(\overline{r_hr_\ell},s_i)$.
Thus $M_2\le\rightI(\overline{r_hr_\ell},s_i)$, the recurrence is in its first
case, and
$\continuousPO_2=\max\{M_2,\leftI(\overline{r_hr_\ell},s_i)\}\le t$.

For~(ii), assume $\continuousPO_2(s_i,S',\overline{r_hr_\ell},R')=t<\infty$.
Then $M_2\le\rightI(\overline{r_hr_\ell},s_i)$, so the interval of~(F1) is
non-empty and $t=\max\{M_2,\leftI(\overline{r_hr_\ell},s_i)\}$ lies in it; in
particular the point $q=(|S'|,|R'|-1+t)$ is free.
Let $s_j$ and the term attaining $M_2$ be fixed.
If $M_2=D_2$, the
induction hypothesis gives a pair
$(\pi',\rho)\in\mathcal P(s_j,S'\setminus\{s_i\},\overline{r_hr_\ell},R')$ for
which $p=(|S'|-1,|R'|-1+M_2)$ is reachable.
If $M_2=\zeta(D_3)=0$,
the induction hypothesis gives a pair
$(\pi,\rho')\in\mathcal P(\overline{s_js_i},S',r_h,R'\setminus\{r_\ell\})$ for
which $p=(|S'|-1+u,|R'|-1)$ is reachable, where
$u=D_3$.
If $M_2=D_1=0$, the
induction hypothesis gives a pair $(\pi',\rho')$ of linear extensions of
$S'\setminus\{s_i\}$ and $R'\setminus\{r_\ell\}$ ending in $s_j$ and $r_h$ for
which $p=(|S'|-1,|R'|-1)$ is reachable.
In each case, appending $s_i$ resp.\ $r_\ell$ where necessary yields a pair
$(\pi,\rho)\in\mathcal P(s_i,S',\overline{r_hr_\ell},R')$, and by the prefix
property $p$ is still reachable in $F_\varepsilon(\pi,\rho)$.
Both $p$ and $q$ lie in the cell $C$, both are free, and $q$ dominates $p$
coordinatewise: in the first case because $M_2\le t$, in the other two cases
because $u\le 1$ and $t\ge 0$.
By~(F2) the straight segment from $p$ to $q$ is a monotone path inside the free
space, so $q$ is reachable.

\emph{Induction step for states of type~(3).}
Symmetric to type~(2), with the roles of the two sides exchanged; here the
relevant coordinate is the one along $\overline{s_js_i}$, which is preserved when
entering $C$ through its bottom edge and is reset to $0$ when entering through
its left edge or its bottom left corner.

\emph{Induction step for states of type~(1).}
For~(i), let $(\pi,\rho)\in\mathcal P(s_i,S',r_\ell,R')$ and let $\gamma$ be a
monotone path to $q=(|S'|,|R'|)$; in particular $q$ is free, that is,
$d(s_i,r_\ell)\le\varepsilon$.
Let $s_j$ and $r_h$ be the points preceding $s_i$ in $\pi$ and $r_\ell$ in
$\rho$; they are maximal in $S'\setminus\{s_i\}$ and $R'\setminus\{r_\ell\}$, so
$(s_j,r_h)$ is one of the pairs over which $M_1$ minimizes.
Exactly as above, the first point $\gamma(\tau)$ of $\gamma$ inside $C$ lies on
the left edge, on the bottom edge, or is the bottom left corner of $C$, and in
these three cases the induction hypothesis shows that
$D_2$, $D_3$ or $D_1$ is finite, respectively.
Hence $M_1\neq\infty$ and $\continuousPO_1(s_i,S',r_\ell,R')=0$.
For~(ii), assume $\continuousPO_1(s_i,S',r_\ell,R')=0$.
Then $d(s_i,r_\ell)\le\varepsilon$, so $q$ is free, and one of the three terms of
$M_1$ is finite for some pair $(s_j,r_h)$.
As in case~(ii) for type~(2), that term provides a pair of linear extensions and
a reachable point $p\in C$, and appending $s_i$ resp.\ $r_\ell$ where necessary
yields a pair $(\pi,\rho)\in\mathcal P(s_i,S',r_\ell,R')$ for which $p$ is still
reachable.
Since $q$ is the top right corner of $C$, it dominates $p$ coordinatewise, so
by~(F2) it is reachable.
This completes the induction.

Finally, $\delta_F(S,R)\le\varepsilon$ holds if and only if $(n,m)$ is reachable
in $F_\varepsilon(\pi,\rho)$ for some linear extensions $\pi$ of $\prec_S$ and
$\rho$ of $\prec_R$.
Every such $\pi$ ends in a maximal element of $S$ and every such $\rho$ ends in a
maximal element of $R$, and conversely every $s_i\in\operatorname{Max}(S)$ and
$r_\ell\in\operatorname{Max}(R)$ occurs as such a last element.
Hence this is the case if and only if $\continuousPO_1(s_i,S,r_\ell,R)=0$ for
some $s_i\in\operatorname{Max}(S)$ and $r_\ell\in\operatorname{Max}(R)$.

\textbf{Complexity.}
Every recursive dependency removes a point from one of the two sets, so the
tables can be filled in increasing order of $|S'|+|R'|$.
The number of states of type~(1) is
$\sum_{S'\in\mathcal J(S)}\sum_{R'\in\mathcal J(R)}
|\operatorname{Max}(S')|\,|\operatorname{Max}(R')|=\mathcal{O}(nm2^{n+m})$.
A state of type~(2) is a state of type~(1) together with the point $r_h$, and a
state of type~(3) is a state of type~(1) together with the point $s_j$, so there
are $\mathcal{O}(nm^22^{n+m})$ states of type~(2) and $\mathcal{O}(n^2m2^{n+m})$
states of type~(3).
Altogether the tables have $\mathcal{O}(nm(n+m)2^{n+m})$ entries, which bounds
the space.

Each entry of type~(1) is evaluated by minimizing over the $\mathcal{O}(nm)$
pairs $(s_j,r_h)$, using three table lookups and a comparison per pair, hence in
$\mathcal{O}(nm)$ time.
Each entry of type~(2) is evaluated by minimizing over the $\mathcal{O}(n)$
choices of $s_j$ in $\mathcal{O}(n)$ time, and each entry of type~(3) in
$\mathcal{O}(m)$ time.
Summing over all states gives
$$
\mathcal{O}\bigl(nm\cdot nm2^{n+m}\bigr)
+\mathcal{O}\bigl(n\cdot nm^22^{n+m}\bigr)
+\mathcal{O}\bigl(m\cdot n^2m2^{n+m}\bigr)
=\mathcal{O}\bigl(n^2m^22^{n+m}\bigr).
$$

It remains to bound the preprocessing.
The lower sets of $S$ and $R$ together with their maximal elements are computed
by a dynamic programming algorithm over bit masks in $\mathcal{O}(n2^n+m2^m)$ operations, as
in the previous subsection.
The intervals of~(F1), and hence all values of $\leftI$ and $\rightI$, are
obtained by solving one quadratic equation each, which takes
$\mathcal{O}(nm^2+n^2m)$ time in total.
The values $d_{\max}(s_i,X)$ for all $X\subseteq R$ and $d_{\max}(r_\ell,Y)$ for
all $Y\subseteq S$ are computed by the recursion
$d_{\max}(p,X)=\max\{d_{\max}(p,X\setminus\{x\}),d(p,x)\}$ in
$\mathcal{O}(n2^m+m2^n)$ time.
All of this is dominated by $\mathcal{O}(n^2m^22^{n+m})$.
\end{proof}

\subsection{Proof of Theorem~\ref{thm:Algo3exact}}
\label{app:proofAlgo3exact}

\begin{proof}
Alt and Godau~\cite{AltGodau1995} showed that for two fixed polygonal curves $P$
and $Q$ the value $\delta_F(P,Q)$ is one of the following critical values:
\begin{enumerate}
    \item[(a)] the distance between the two start points or the distance between
    the two endpoints;
    \item[(b)] the distance between a vertex of one curve and an edge of the
    other, that is, the smallest $\varepsilon$ for which a free interval on a
    cell boundary becomes non-empty;
    \item[(c)] the common distance of two vertices of one curve to a point on an
    edge of the other, where this point is the intersection of the edge with the
    perpendicular bisector of the two vertices, that is, the smallest
    $\varepsilon$ for which two free intervals on one cell boundary line merge
    and a monotone passage opens.
\end{enumerate}
Every curve induced by $S$ has its vertices in $S$ and its edges among the
segments spanned by two points of $S$, and analogously for $R$.
Hence for all linear extensions $\pi$ of $\prec_S$ and $\rho$ of $\prec_R$, the
critical values of the pair $(c_\pi(S),c_\rho(R))$ are contained in
$\mathcal C=\mathcal C_a\cup\mathcal C_b\cup\mathcal C_c$, where $s,s'$ range
over $S$, $r,r'$ range over $R$, $b(p,q)$ denotes the perpendicular bisector of
$p$ and $q$, and
\begin{align*}
\mathcal C_a&=\{d(s,r)\},
    \qquad
    \mathcal C_b=\{d(s,\overline{rr'})\}\cup\{d(r,\overline{ss'})\},\\
\mathcal C_c&=\{d(x,r)\mid x\in\overline{ss'}\cap b(r,r')\}
    \cup\{d(y,s)\mid y\in\overline{rr'}\cap b(s,s')\}.
\end{align*}
Note that $|\mathcal C_a|=\mathcal{O}(nm)$,
$|\mathcal C_b|=\mathcal{O}(nm^2+n^2m)$ and
$|\mathcal C_c|=\mathcal{O}(n^2m^2)$, so $|\mathcal C|=\mathcal{O}(n^2m^2)$, and
$\mathcal C$ can be computed in $\mathcal{O}(n^2m^2)$ time, since each of its
elements is determined in constant time by at most two points of each input set.

There are finitely many pairs of linear extensions, so the minimum in
$\delta_F(S,R)=\min_{\pi,\rho}\delta_F(c_\pi(S),c_\rho(R))$ is attained, and by
the above $\delta_F(c_\pi(S),c_\rho(R))\in\mathcal C$ for every pair
$(\pi,\rho)$.
Hence $\delta_F(S,R)\in\mathcal C$.
Moreover, the predicate $\delta_F(S,R)\le\varepsilon$ is monotone in
$\varepsilon$, and by Theorem~\ref{thm:Algo3} it can be decided in
$\mathcal{O}(n^2m^22^{n+m})$ time for every fixed $\varepsilon$.
Therefore the smallest $\varepsilon\in\mathcal C$ for which the decision
procedure answers yes equals $\delta_F(S,R)$, and it can be found by binary
search over the sorted sequence of the values in $\mathcal C$ using
$\mathcal{O}(\log|\mathcal C|)=\mathcal{O}(\log(nm))$ calls of the decision
procedure.
Computing and sorting $\mathcal C$ takes $\mathcal{O}(n^2m^2\log(nm))$ time and
is dominated by a single call of the decision procedure.
The total running time is thus $\mathcal{O}(n^2m^22^{n+m}\log(nm))$.
\end{proof}

\subsection{Proof of Proposition~\ref{prop:eth} and Corollary~\ref{cor:sh}}
\label{app:eth}
 
We compose three reductions, each of which increases the size of the instance by
at most a constant factor.
Throughout this section, a 3-SAT formula has $v$ variables and $c$ clauses, a
NAE-3SAT formula has $\mu$ clauses, a $(3,B2)$-SAT formula has $V$ variables and
$C$ clauses, and $n=|S|$, $m=|R|$ as in the rest of the paper.
 
\paragraph{Step 1: 3-SAT to NAE-3SAT.}
Let $\varphi$ be a 3-SAT formula with $v$ variables and $c$ clauses.
Introduce one new variable $s$ and replace every clause $(l_1\vee l_2\vee l_3)$
of $\varphi$ by the not-all-equal clause $\mathrm{NAE}(l_1,l_2,l_3,s)$.
Since the set of satisfying assignments of a NAE formula is closed under
complementation, we may assume $s$ to be false in a satisfying assignment, and
then $\mathrm{NAE}(l_1,l_2,l_3,s)$ holds if and only if one of $l_1,l_2,l_3$ is
true.
Hence $\varphi$ is satisfiable if and only if the resulting NAE formula is.
Each clause of size four is then split by introducing a fresh variable $z$ and
replacing $\mathrm{NAE}(a,b,c,d)$ by
$\mathrm{NAE}(a,b,z)\wedge\mathrm{NAE}(\neg z,c,d)$, which preserves
satisfiability: if $a\ne b$ or $c\ne d$ a suitable $z$ exists, and if
$a=b=c=d$ the first clause forces $z\ne a$ and the second becomes
$\mathrm{NAE}(a,a,a)$.
The result is a NAE-3SAT formula with $\mu=2c$ clauses, all of size exactly
three, and $v+1+c$ variables.
 
\paragraph{Step 2: NAE-3SAT to $(3,B2)$-SAT.}
We use the construction of Berman, Karpinski and
Scott~\cite[Section~3]{BermanKarpinskiScott2004} and count its size.
Recall their enforcer: for literals $l_0,l_1,l_2$ the formula
$$
S(l_0,l_1,l_2)\ \equiv\
(l_0\vee\neg a\vee b)\wedge
(l_1\vee\neg b\vee c)\wedge
(l_2\vee a\vee\neg c)\wedge
(a\vee b\vee c)\wedge
(\neg a\vee\neg b\vee\neg c)
$$
with new variables $a,b,c$ is satisfiable in $a,b,c$ if and only if
$l_0\vee l_1\vee l_2$ holds, it consists of five clauses, each of
$a,b,c$ occurs twice positively and twice negatively in it, and each of
$l_0,l_1,l_2$ occurs once.
For a literal $l$, the enforcer $l^{(2)}=S(l,y,y)\cup S(l,\neg y,\neg y)$ with a
new variable $y$ forces $l$ to be true, consists of ten clauses and seven new
variables, and contains the literal $l$ exactly twice.
The reduction now proceeds in five stages.
\begin{enumerate}
    \item Replace every occurrence of a variable by a new variable.
    A NAE-3SAT formula with $\mu$ clauses thus gets $3\mu$ variables, each
    occurring exactly once.
    \item Replace every clause $\mathrm{NAE}(l_1,l_2,l_3)$ by
    $(l_1\vee l_2\vee l_3)\wedge(\neg l_1\vee\neg l_2\vee\neg l_3)$.
    This gives $2\mu$ clauses, and every variable now occurs exactly once
    positively and once negatively.
    \item For the variables $x_0,\dots,x_k$ that replaced one variable of the
    original formula, add the wheel of implications
    $x_0\to x_1,\ \dots,\ x_k\to x_0$.
    Over all original variables this produces one implication per occurrence,
    that is $3\mu$ implications in total, and every $x_i$ gains one positive and
    one negative occurrence.
    \item Replace every implication $x\to y$ by the clause
    $(\neg x\vee y\vee\neg a)$, where the implications are paired arbitrarily and
    the two implications of a pair share the same new variable $a$.
    In our chain $3\mu=6c$ is even, so a perfect pairing exists.
    This gives $3\mu$ clauses and $\tfrac32\mu$ new variables, and the literal
    $\neg a$ occurs exactly twice for every new variable $a$.
    \item For every new variable $a$ of the previous stage add the enforcer
    $a^{(2)}$, which forces $a$ to be true, so that the clauses of stage~4 are
    equivalent to the implications.
    This gives $10\cdot\tfrac32\mu=15\mu$ clauses and
    $7\cdot\tfrac32\mu=\tfrac{21}2\mu$ new variables, and supplies the two
    missing occurrences of the literal $a$.
\end{enumerate}
The resulting formula $g$ is satisfiable if and only if the NAE-3SAT formula is,
and it has
$$
C=2\mu+3\mu+15\mu=20\mu
\quad\text{clauses and}\quad
V=3\mu+\tfrac32\mu+\tfrac{21}2\mu=15\mu
\quad\text{variables}.
$$
Every literal occurs exactly twice, which is confirmed by
$3C=60\mu=4V$.
With $\mu=2c$ from Step~1 we obtain a $(3,B2)$-SAT formula with $V=30c$
variables and $C=40c$ clauses.
 
\paragraph{Step 3: $(3,B2)$-SAT to point sets.}
We use the construction in the proof of
\cite[Theorem~2]{BUCHIN2022}, which for a $(3,B2)$-SAT formula $A$ and a
constant $\hat c>1$ builds point sets $S,R\subset\mathbb R^2$ and a value
$\varepsilon>0$ in polynomial time such that $A$ is satisfiable if and only if
$\delta_F(S,R)\le\varepsilon$.
Its points are, per clause of $A$, one clause point of $R$, and, per variable
$x$ of $A$, the four variable-assignment points $s_{x_1},s_{x_2}$ and
$s_{\bar x_1},s_{\bar x_2}$ of $S$, the four variable-assignment points
$r_{x_1},r_{x_2}$ and $r_{\bar x_1},r_{\bar x_2}$ of $R$, and the two
segment-forced points $r_{(x\bar x)_1},r_{(x\bar x)_2}$ of $R$.
Hence $|S|=4V$ and $|R|=6V+C$.
Using $C=\tfrac43V$ this gives
$$
n+m=|S|+|R|=4V+6V+\tfrac43V=\tfrac{34}3V .
$$
With $V=30c$ from Step~2 we obtain $n+m=340c$.
 
\paragraph{Composition.}
Given a 3-SAT formula $\varphi$ with $c$ clauses, Steps~1 to~3 construct in
polynomial time two point sets $S,R\subset\mathbb R^2$ and a value
$\varepsilon>0$ with $n+m=340c$ such that $\varphi$ is satisfiable if and only if
$\delta_F(S,R)\le\varepsilon$; with the normalization of the first remark below
the constant $340$ grows by a constant factor. This proves
Proposition~\ref{prop:eth}. Consequently, if $\delta_F(S,R)\le\varepsilon$ can be
decided in time $T(n+m)$ for a non-decreasing function $T$, then 3-SAT on formulas
with $c$ clauses can be decided in time $T(\kappa c)+\operatorname{poly}(c)$ for a
constant $\kappa$.

\paragraph{Proof of Corollary~\ref{cor:sh}.}
Let $\psi$ be a CNF formula of length $L$. If $\psi$ contains an empty clause, it
is unsatisfiable. Otherwise, replacing every clause $(l_1\vee\dots\vee l_w)$ with
$w>3$ by the clauses $(l_1\vee l_2\vee y_1)$, $(\neg y_1\vee l_3\vee y_2)$,
\dots, $(\neg y_{w-3}\vee l_{w-1}\vee l_w)$ with new variables
$y_1,\dots,y_{w-3}$, and padding every clause with fewer than three literals by
repeating one of its literals, gives an equisatisfiable 3-SAT formula with $c\le L$
clauses. Let $\gamma>0$; we may assume $\gamma<1$, since the statement for
$\gamma\ge 1$ follows from the one for $\gamma=\frac12$. Suppose that $\delta_F(S,R)\le\varepsilon$ can be decided in time
$2^{a(n+m)^{1-\gamma}}B^a$ for a constant $a$, where $B$ is the bit length of the
input. With the rational version of the construction described in the second
remark below, the composition above transforms $\psi$ in polynomial time into an
instance with $n+m\le\kappa L$ and $B=\mathcal O(L\log L)$, so the satisfiability
of $\psi$ can be decided in time
$2^{a(\kappa L)^{1-\gamma}}\operatorname{poly}(L)$. Since
$a\kappa^{1-\gamma}L^{1-\gamma}+\mathcal O(\log L)\le\frac12L^{1-\gamma/2}$ for all
sufficiently large $L$, this is $\mathcal O(2^{L^{1-\gamma/2}})$, which
contradicts SH. 
For partially
ordered point sets the statement follows by equipping $S$ and $R$ with the empty
partial orders, in which case every ordering of the points is a linear extension
and the two distances coincide. In particular, the exponent $n+m$ in the running
time $\mathcal O(n^2m^22^{n+m})$ of Theorem~\ref{thm:Algo3} cannot be improved to
$(n+m)^{1-\gamma}$, and neither can the exponent in Theorem~\ref{thm:Algo3exact},
since computing $\delta_F(S,R)$ decides $\delta_F(S,R)\le\varepsilon$.
 
\paragraph{Two remarks on the construction.}
First, the construction of Step~3 assumes that no two clauses of $A$ have two
literals in common.
This assumption is local: since every literal occurs exactly twice, a pair of
literals is contained in at most two clauses, so at most $3C$ pairs of clauses
violate it, and repairing one of them affects a constant number of clauses and
variables.
The normalization therefore increases the size of $A$ by at most a constant
factor and does not affect the argument.
Second, in~\cite{BUCHIN2022} the clause points are placed on consecutive corners
of a regular polygon and $\varepsilon$ is a fraction of a minimum point-segment
distance, so in general neither the coordinates nor $\varepsilon$ are rational.
Since SH is a statement about Turing machines, we use the following rational
version of the construction. The clause points are only required to be in
strictly convex position on a circular arc of angle less than $\pi/2$; we place
the $k$-th of the $C$ clause points at the rational point
$\bigl(\frac{1-t_k^2}{1+t_k^2},\frac{2t_k}{1+t_k^2}\bigr)$ of the unit circle with
$t_k=\frac{k}{C+1}$. Every variable point lies on the line through two clause
points $p$ and $q$, on a prescribed side of them, and has to avoid a polynomial
number of forbidden positions on this line; we choose it as $p+\lambda(q-p)$
with $\lambda$ among polynomially many consecutive integers, at least one of which
is allowed. Similarly, every segment-forced point is chosen as $s+\mu(s'-s)$ on a
segment $\overline{ss'}$ between two variable points, with $\mu\in\{\frac
jK\mid 0<j<K\}$ for a polynomial $K$. Hence all coordinates are rational numbers
with $\mathcal O(\log C)$ bits. The proof of~\cite{BUCHIN2022} uses $\varepsilon$
only through $0<\varepsilon<\hat\varepsilon$, where $\hat\varepsilon$ is the minimum
distance of a clause point to a segment between points of different literals and
of a segment-forced point to a segment between points of different variables.
Each such squared distance is a rational number with $\mathcal O(\log C)$ bits, so
$\varepsilon=2^{-k}$ for the smallest integer $k$ with
$2^{-2k}<\hat\varepsilon^{\,2}$ has $\mathcal O(\log C)$ bits as well. The constructed instance therefore has encoding length
$B=\mathcal O((n+m)\log(n+m))$ and is computed in polynomial time on a Turing
machine, as used in the proof of Corollary~\ref{cor:sh} above.

\end{document}